# INTEGRAL OBSERVATIONS OF THE VELA REGION FOCUSING ON VELA X-1

Stéphane Schanne[1,*], Diego Götz[1], Lucie Gérard[1], Patrick Sizun[1],
Maurizio Falanga[1], Clarisse Hamadache[1], Bertand Cordier[1], Andreas von Kienlin[2]

[1] CEA-Saclay, DAPNIA/Service d'Astrophysique, F-91191 Gif sur Yvette, France
[2] Max-Planck-Institut für extraterrestrische Physik, D-85740 Garching, Germany
[*] corresponding author: schanne{at}hep.saclay.cea.fr

## ABSTRACT

The Vela region has been observed for 1.7 Ms in November 2005 by the INTEGRAL satellite. We present preliminary spectral and temporal results of Vela X-1, an eclipsing neutron star hosted in a wind-accreting high-mass X-ray binary system. Using data from ISGRI, SPI and JEM-X, we firmly confirm the existence of cyclotron resonant scattering features (CRSF) at ~27 keV and ~54 keV, implying a neutron-star magnetic field of $3 \times 10^{12}$ Gauss, and the presence of an iron emission line at ~6.5 keV. During two strong flares those parameters remained unchanged. Furthermore we measure the neutron-star spin period of 283.6 s, indicating a still constant trend.

## 1. INTRODUCTION

The Vela region has been observed by INTEGRAL [1], ESA's gamma-ray satellite, in the frame of our accepted AO-3 open-time proposal. Data were taken from November 4 to December 4, 2005, during 1.7 Ms, from INTEGRAL revolution n°373 to 383, in 473 different pointings. The key target is the detection of γ-rays emitted by the decay of radioactive isotopes produced in stellar nucleosynthesis processes as e.g. 1809 keV emission from $^{26}$Al decays. Results of the $^{26}$Al studies are presented in [2]. Besides nucleosynthesis studies, the region is also of interest for the presence of several point sources, among which Vela X-1, for which we present preliminary results in this paper.

In the soft γ-ray band the brightest point source in the region is Vela X-1, an eclipsing wind accreting high-mass X-ray binary system (HMXB) consisting of a neutron star (NS) and a massive, 23 solar masses ($M_\odot$), donor star (HD 77581) classified as a B0.5 super-giant [3]. The system has an orbital period of 8.964 days [4]. The NS is deeply embedded in the intense stellar wind of the companion, whose inferred mass-loss rate is of the order of $10^{-7}$ $M_\odot$/yr [5]. The NS shows a spin period of ~283 s in X-rays (e.g. [6]).

## 2. SEARCH FOR SOURCES IN VELA REGION

Using the full IBIS/ISGRI data set, we search for point sources in the Vela region, by constructing a mosaic image of the region (Fig. 1) in the 18-60 keV energy range, most suited for new source detection. The sources found comprise four sources (Tab. 1) –all of them known– which are detected with significance above 6 σ, considered to be the limit for firm source detection.

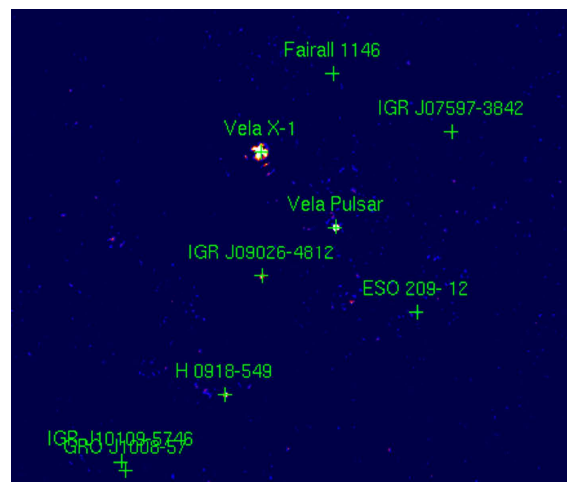

*Fig. 1. Image of the Vela region by ISGRI (18-60 keV).*

| Source | σ-level |
|---|---|
| Vela X-1 | 1316.3 |
| Vela Pulsar | 48.2 |
| H 0918-548 | 13.9 |
| IGR J09026-4812 | 12.0 |
| IGR J07597-3842 | 5.77 |
| GRO J1008-57 | 5.46 |
| Fairall 1146 | 5.45 |
| ESO 209-12 | 5.11 |
| IGR J10109-5746 | 5.08 |

*Tab. 1. Sources found by ISGRI in the Vela region (18-60 keV energy band, 1.7 Ms exposure). Only the first four sources can be considered firmly detected.*

A similar search has been performed with the SPI data. We use SPIROS imaging in the 20-40 keV energy band. The net exposure time was 1 Ms after selecting the best 271 pointings ($\chi^2$<10 in SPIROS). The SPIROS mode used was a blind search for 3 sources. The background model used empty fields close in time to the Vela data taking period, however the saturating Ge-detector background model yields similar results. We obtain a Vela region image (Fig. 2) and the results are summarized in Tab. 2. Two sources, Vela X-1 and Vela Pulsar, are firmly detected, while the 3rd source might be associated with IGR J07597-3842 or IGR J07565-4139, both located within the SPI angular resolution at a distance<2° from the detected position.

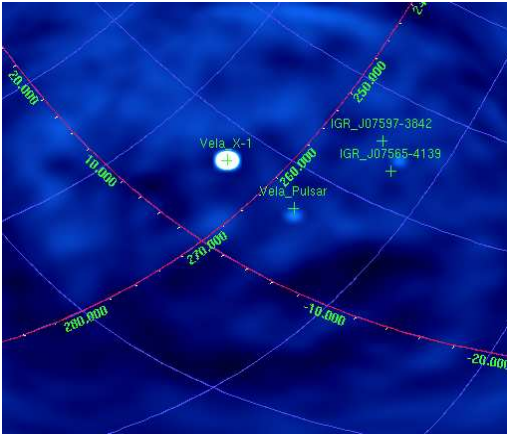

*Fig. 2. SPIROS significance image of the Vela region (20-40 keV) obtained in a blind search for three sources.*

| Source | Sigma | Cat Source | Distance |
|---|---|---|---|
| 1 | 235.1 | Vela X1 | 0° |
| 2 | 11.6 | Vela Pulsar | ~0.5° |
| 3 | 9.7 | IGR J07597-3842 | ~1° |
|   |     | IGR J07565-4139 | ~2° |

*Tab. 2. Sources found by SPIROS (20-40 keV) in a blind search for three sources. The distances of those sources to known sources in the region are listed.*

## 3. VELA X-1 LIGHTCURVES

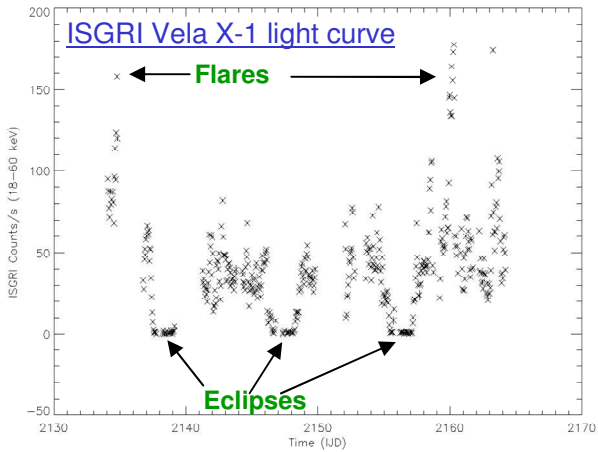

*Fig. 3. ISGRI light curve of Vela X-1. Energy 18-60 keV, time bin 1h, time measured in days since January 1st, 2000 (IJD).*

The Vela X-1 light curves, observed respectively with ISGRI and SPI are shown in Fig. 3 and Fig. 4. Both instruments detect the same Vela X-1 variability, typical for wind accreting pulsars, namely three NS eclipses by the companion star, as well as two strong flares (one in revolution 373 and one in revolution 382).

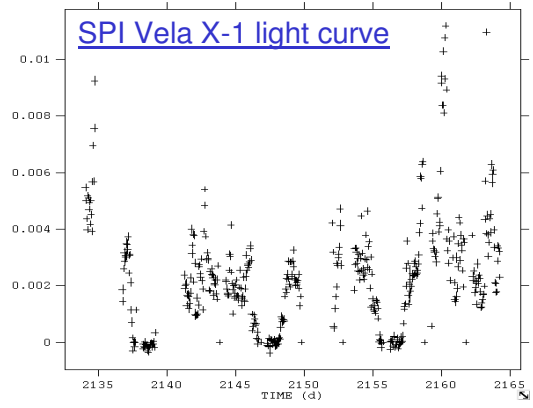

*Fig. 4. SPI light curve of Vela X-1 (20-40 keV) in ph/cm$^2$/s.*

## 4. VELA X-1 FOLDED LIGHTCURVE

Using the barycenter-corrected ISGRI light-curves, spanning over a single pointing (science window) of ~1 h duration, we find the NS spin period to be 283.6±0.1 s, consistent with previous findings [13]. The pulse profiles in two ISGRI energy bands (20-40 keV and 40-100 keV) are shown in Fig. 5. A deeper analysis is ongoing.

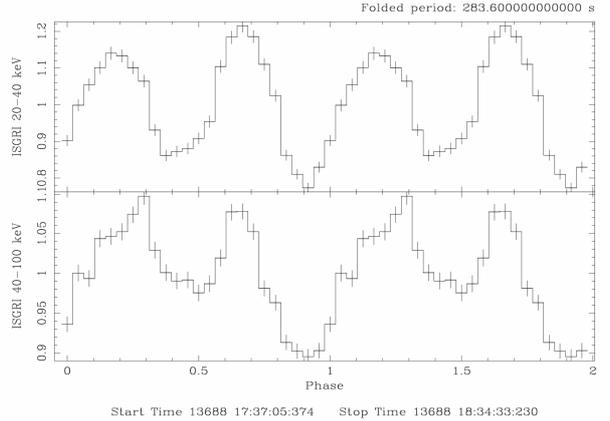

*Fig. 5. ISGRI light curve of Vela X-1 folded at the best period of 283.6 s. Two periods are shown for clarity.*

## 5. VELA X-1 PERSISTENT AVERAGE SEPCTRA

The presence of two cyclotron resonant scattering features (CRSF) in the average spectrum of Vela X-1 has been debated since a long time in the literature [7-15]. With our analysis we assess the presence of two CRSF with fundamental energy at ~27 keV and the 1st harmonic at ~54 keV in agreement with the RXTE observations by Kreykenbohm et al. (2002) [11], and at variance with the single CRSF at ~55 keV reported by La Barbera et al. (2003) [12] using BeppoSAX data. We confirm the real nature of the debated ~27 keV feature, seen independently by the two different instruments ISGRI and SPI.

We derive the persistent average spectrum of Vela X-1, using JEM-X and ISGRI, by removing the time intervals of the two strong flares and the eclipses (80 ks exposure available for JEM-X, 737 ks for IBIS). The fit with a simple cutoff power law (Fig. 6) displays the presence in the data of residuals at ~6.5 keV, ~27 keV and ~54 keV. We model the low-energy excess with a Gaussian emission line, and describe the two high-energy dips with the cyclotron absorption model cyclabs of XSPEC [16]. This yields a fit with a reduced $\chi^2$ of 1.2 applying 2% systematics (Fig. 7). We associate the Gaussian emission feature found at 6.44 keV with a Fe emission line. The low-energy cyclotron absorption line is found to be located at 27.0±0.3 keV. Note that the ISGRI/JEM-X inter-calibration factor is around 1.3. Including the SPI data (Fig. 8), the same model yields a fit with a reduced $\chi^2$ of 1.2 applying 3% systematics, a Fe emission line at 6.56 keV and the low energy cyclotron line located at 27.2 keV.

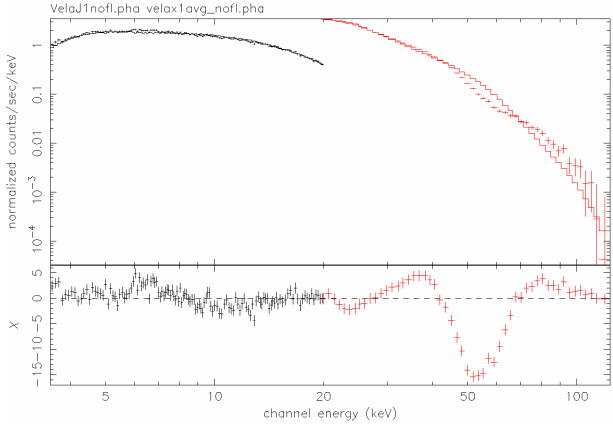

*Fig. 6. INTEGRAL JEM-X and ISRGI persistent average spectrum of Vela X-1, fit with a simple cutoff power law. The upper panel shows the data and the best fit model (solid line); the lower panel shows the residuals.*

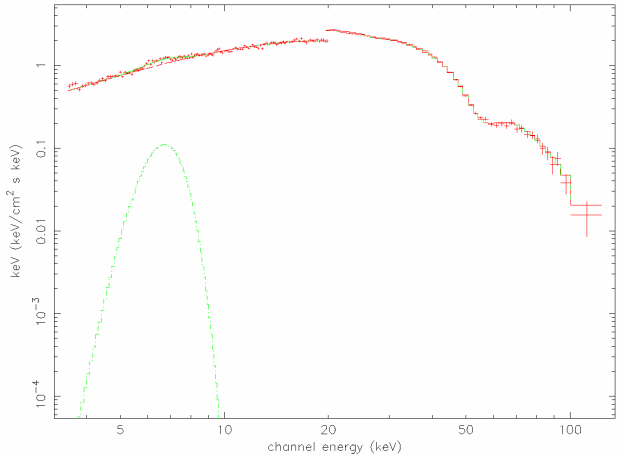

*Fig. 7. INTEGRAL JEM-X and ISRGI unfolded persistent average spectrum of Vela X-1.*

The Vela X-1 persistent average SPI spectrum using SPIROS is shown in Fig. 9. This spectrum is modeled using constant, cutoff power-law, two CRSF features with energies fixed by $E_1=E_2/2$ and fluxes modeled by:

$$F_{CRSF}(E) = \exp\left[-\frac{D(WE/E_i)^2}{(E-E_i)^2 + W^2}\right]$$

The result of the fit is shown in Tab. 3 (1st data column). The fit using the same model applied to both SPI and ISGRI data, leads to compatible results, shown in Tab. 3 (2nd data column).

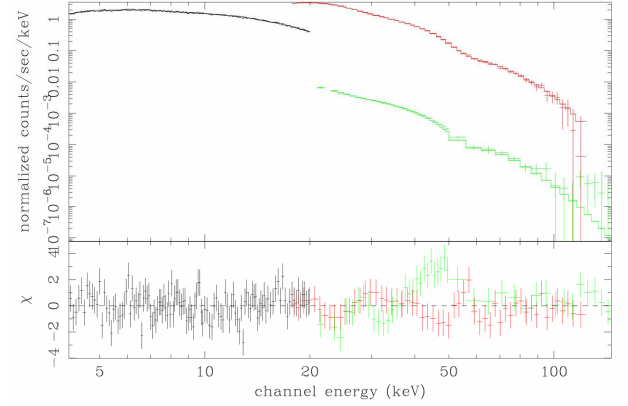

*Fig. 8. JEM-X, ISRGI and SPI persistent average spectrum of Vela X-1 (combined fit).*

|  | Persist. avr. SPI | Persist. avr. SPI + ISGRI | In flares SPI |
|---|---|---|---|
| $E_1=E_2/2$ | 27.44 ± 0.43 | 27.88 ± 0.14 | 27.5 ± 0.6 |
| $W_1$ | 6.73 ± 1.6 | 12.09 ± 0.94 | 5.58 ± 1.69 |
| $D_1$ | 0.17 ± 0.052 | 0.28 ± 0.078 | 0.16 ± 0.049 |
| $W_2$ | 8.84 ± 1.96 | 10.86 ± 1.05 | 8.45 ± 2.96 |
| $D_2$ | 1.01 ± 0.12 | 1.1 ± 0.07 | 0.76 ± 0.15 |
| $\chi^2$ | 1.03 | 1.9 | 1.17 |
| ndf | 71 | 90 | 71 |

*Tab.3. Results of fits to the SPI and SPI+ISGRI persistent average spectra, as well as the SPI spectrum during flares, using the model "constant, cutoff power-law, 2 CRSF (with fixed energy ratios)".*

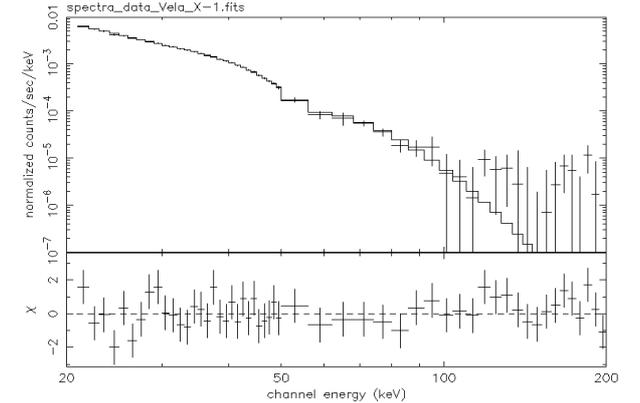

*Fig. 9. SPI persistent average spectrum of Vela X-1.*

## 6. VELA X-1 SEPCTRA DURING FLARES

For the Vela X-1 flare in revolution 373, one pointing is used to derive the combined JEM-X and ISGRI spectrum, shown in Fig. 10. In this limited dataset, the presence of the Fe emission line is still significant at ~6.5 keV, and the low-energy CRSF feature (fit with cyclabs) is found at 26.0±0.4 keV. For the flare in revolution 382, nine IBIS pointings are used, and the low-energy CRSF feature is found at 27.0±0.3 keV.

The SPI spectrum obtained by combining the flares of Vela X-1 in revolutions 373 and 382 is presented in Fig. 11. The result of the fit using the previously presented model with two CRSF is reported in Tab. 3 (3rd data column). We compare those parameters with the ones found for the persistent average spectrum using either SPI alone or SPI+ISGRI data. We conclude that no important change in the structure of the CRSF features can be noticed for those Vela X-1 flares.

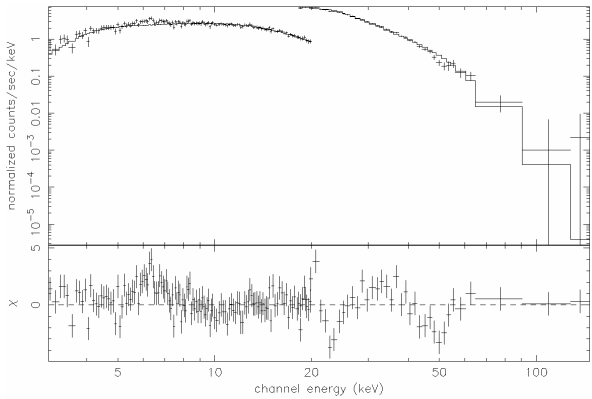

Fig. 10. JEM-X and ISRGI spectrum of Vela X-1 during flare in rev. 373, fitted with a simple cutoff power law, showing the presence of residuals at ~6.5 keV, ~26 keV and ~52 keV.

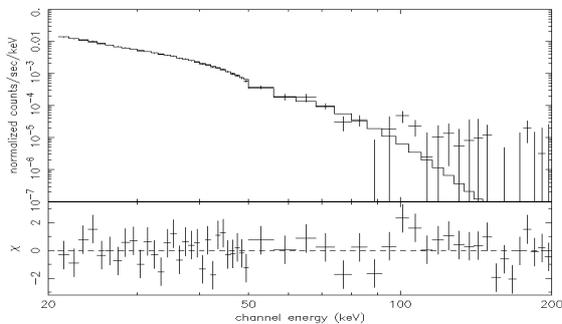

Fig. 11. SPI spectrum of Vela X-1 during flare, fitted with a cutoff power-law and two CRSF.

## 7. CONCLUSIONS

The INTEGRAL satellite observed the Vela region in the X-ray and gamma-ray energy bands during 1.7 Ms between 4th November and 4th December 2005, from revolution 373 to 383, in the context of our accepted AO-3 INTEGRAL open-time observation. We present preliminary results of the analysis of the ISGRI, SPI and JEM-X data. We focus on a spectral and temporal study of the eclipsing high-mass X-ray binary Vela X-1. Thanks to this long-term observation we derive the time-averaged spectrum of the source and firmly confirm the existence of cyclotron resonant scattering features (CRSF) at ~27 keV and ~54 keV using IBIS/ISGRI and SPI data separately and in a combined fit. Assuming canonical parameters for the neutron star (mass 1.4 $M_\odot$, radius 10 km), a 27 keV CRSF implies a neutron star magnetic field of $3 \times 10^{12}$ Gauss. Using the JEM-X data we also found the presence of a Fe emission line at ~6.5 keV. During two strong Vela X-1 flares we found that the energy positions of the CRSF features remain unchanged with respect to their persistent average values. Furthermore we obtain a spin period of 283.6 s by folding the Vela X-1 light curve over one pointing of 1 h duration. In order to complete the Vela X-1 analysis, work is currently ongoing. We will derive a precise orbital ephemeris, which is important for the study of this system, since it allows to determine detailed time-resolved spectra during the eclipse ingress and egress phases in different energy bands, and to monitor details of the companion wind. Orbital and pulse-phase resolved spectroscopy will permit us to understand the underlying physics of the high-energy emission. Using a relativistic description of the polar-cap emission of the neutron star, the variation of the pulse morphology with energy and time will permit to model the pulsed flux in energy and constrain the system geometry.